# Quantum paramagnetism in the hyperhoneycomb Kitaev magnet β-ZnIrO$_3$

Yuya Haraguchi[1,†], Akira Matsuo[2], Koichi Kindo[2], and Hiroko Aruga Katori[1,3]
[1]Department of Applied Physics and Chemical Engineering, Tokyo University of Agriculture and Technology, Koganei, Tokyo 184-8588, Japan
[2]The Institute for Solid State Physics, The University of Tokyo, Kashiwa, Chiba 277-8581, Japan
[3]Research Center for Thermal and Entropic Science, Graduate School of Science, Osaka University, Toyonaka, Osaka 560-0043, Japan
†chiyuya3@go.tuat.ac.jp

A polycrystalline sample of the hyperhoneycomb iridate β-ZnIrO$_3$ was synthesized via a topochemical reaction, and its structural, magnetic, and thermodynamic properties were investigated. The magnetization and heat capacity data show the absence of long-range magnetic order at least down to 2 K. A positive Curie-Weiss temperature $\theta_W \sim$ 45 K probed by the temperature dependence of magnetic susceptibility indicates that a Kitaev interaction is dominant. These observations suggest that a quantum spin liquid may have been realized. Furthermore, the observation of linear temperature-dependent contribution to the heat capacity with no magnetic field effect evidences gapless excitation. These facts are surprisingly contrary to the chemical disorder evidenced by the crystallographic analysis. By discussing the differences in the size effect of $Z_2$-fluxes in 2D and 3D Kitaev magnets, we propose that there is a hidden mechanism to protect the quantum spin liquid state from chemical disorder.

*Introduction*. The Kitaev model—a model recognized by Ising-like bond-dependent interactions on a tricoordinated lattice—has attracted much attention as the most promising platform to realize a quantum spin liquid state (QSL) [1-4]. One of the essential features in the Kitaev QSL is that the spins are fractionalized into itinerant Majorana fermions and localized $Z_2$-fluxes [1,5-8]. The Kitaev model is materialized by the superexchange interactions between spin-orbital-coupled $J_{\text{eff}} = 1/2$ electrons in 4$d$/5$d$ transition metal ions with a low-spin $d^5$ electron configuration through the Jackeli-Khaliullin mechanism [2].

In realistic compounds, the Kitaev QSL state is easily destabilized by the unquenched non-Kitaev interactions, for example, the Heisenberg interaction $J$ and the off-diagonal interaction $\Gamma$ [9-12]. Indeed, almost all Kitaev candidate materials exhibit long-range magnetic orderings. The magnitude of these interactions is strongly dependent on the microscopic environment around Ir ions, especially the Ir-O-Ir bridging angle and trigonal distortion of IrO$_6$ octahedra [13,14].

Under the circumstance, one can consider that the Kitaev interaction term $K$ can be relatively enhanced by reducing the $J$ and $\Gamma$ terms with appropriate adjustments of the crystal structure. There are two feasible and effective tuning strategies. One is the application of hydrostatic pressure. At first glance, it appears that pressure can be applied to adjust the lattice parameters without disturbing the system. However, all of these attempts have failed so far—previous high-pressure studies have shown that the formation of molecular orbitals due to increased orbital overlap between ions causes them to become nonmagnetic in α-Li$_2$IrO$_3$, Na$_2$IrO$_3$ and α-RuCl$_3$ [15-18].

The other strategy is the topochemical manipulations. Li-ions in α-Li$_2$IrO$_3$ or Na-ions in Na$_2$IrO$_3$ are wholly or partially replaced soft-chemically by other monovalent or divalent ions [19]. This manipulation tunes the entire lattice and produces changes in the ratio of $J$, $K$, and $\Gamma$. Examples include delafossite-type Ag$_3$LiIr$_2$O$_6$ [20], H$_3$LiIr$_2$O$_6$ [21], Cu$_2$IrO$_3$ [22], and ilmenite-type MgIrO$_3$, ZnIrO$_3$, and CdIrO$_3$ [23,24]. All delafossite-type compounds exhibit spin-liquid-like behavior [20-22]. However, the crystallographical disorder has been found to exist in all of them [25-27]. Thus, such a disorder can mimic a spin-liquid-like state without the dominant Kitaev interactions. In contrast, all ilmenite-type iridates exhibit magnetic ordering [23,24]. In addition, the antiferromagnetic interaction is dominant in all of them, likely representing a deviation from the pure Kitaev model.

In parallel with this, three-dimensional (3D) analogs of the two-dimensional (2D) honeycomb of Ir$^{4+}$ ions were discovered as another platform for Kitaev magnetism [28-30]. Examples include the "hyperhoneycomb" lattice in β-Li$_2$IrO$_3$ [28, 29] and the "harmonic honeycomb" lattice in γ-Li$_2$IrO$_3$ [30] defined on several 3D tricoordinate lattices. The magnetic susceptibility of both compounds evidences the predominant ferromagnetic exchange interactions, likely indicating that the Kitaev-term is dominant. This origin is probably due to the more negligible effect of trigonal strain and stacking defects than the 2D system, originating from its 3D structure. Thus, the lattice tuning with the topochemical manipulation to the three-dimensional Li$_2$IrO$_3$ polymorphs would effectively approach the pure Kitaev model for realizing the Kitaev QSL.

In this letter, we report on the successful synthesis of a hyperhoneycomb iridate β-ZnIrO$_3$ via the topochemical reaction to β-Li$_2$IrO$_3$ as well as its structural, magnetic, and thermodynamic properties. The crystal structure analysis revealed the existence of a random distribution of nonmagnetic Zn ions. Nevertheless, no clear magnetic/glass transition or disorder-induced scaling behavior has been observed. These results indicate that the robustness of Kitaev QSL is strongly related to their dimensionality, which is probably due to the stability of the localized $Z_2$-fluxes differing between 2D and 3D systems [31,32].

*Experimental Procedure.* The precursor β-Li$_2$IrO$_3$ was prepared via the conventional solid-state reactions. Stoichiometric amounts of Li$_2$CO$_3$ and IrO$_2$ were mixed, and the mixture was calcined at 1000°C for 24 h in air and then quenched. This precursor was ground well with a significant excess of ZnSO$_4$ and an inert salt KCl in an Ar-filled glove box, sealed in an evacuated Pyrex tube, and reacted at 400°C for 100 h. The ion-exchange reaction is expressed as

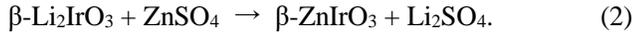

$$\beta\text{-Li}_2\text{IrO}_3 + \text{ZnSO}_4 \rightarrow \beta\text{-ZnIrO}_3 + \text{Li}_2\text{SO}_4. \qquad (2)$$

Initially, the synthesis using ZnCl$_2$ as a common 2Li$^+$↔Zn$^{2+}$ ion-exchange reaction agent was attempted. However, after the reaction with ZnCl$_2$, the Pyrex glass was filled with chlorine gas, suggesting the non-redox-neutral process. Thus-obtained samples show significant sample dependence with a significant spin-glass-like contribution. Therefore, the valence of Ir is expected to be deviated from tetravalent in the sample obtained using ZnCl$_2$. On the other hand, we found that the ion exchange reaction can be substituted using ZnSO$_4$ instead of ZnCl$_2$. This method does not produce chlorine gas naturally, and a redox-neutral reaction is expected. However, when only ZnSO$_4$ was used as the ion-exchange reactant, the reaction rate was prolonged, and it was not easy to obtain a single-phase sample. In this study, we applied the method applied to the previously-reported ion-exchange reaction from ilmenite-type oxide NaSbO$_3$ to rosiaite-type oxide ZnSb$_2$O$_6$ [33]. We succeeded in increasing the reaction rate by adding inert salt of KCl to lower the melting point and thus promote ion diffusion. In addition, the sample dependence of magnetization was negligible (see Supplemental Material [34]). Therefore, the reaction using ZnSO$_4$/KCl mixture salt can be assumed to be a redox-neutral process. The remained salts were removed by washing the sample with distilled water.

The thus-obtained sample was characterized by powder x-ray diffraction (XRD) experiments in a diffractometer with Cu-Kα radiation. The cell parameters and crystal structure were refined by the Rietveld method using the RIETAN-FP v2.16 software [35]. The temperature dependence of dc and ac magnetization was measured under several magnetic fields in a magnetic property measurement system (MPMS; Quantum Design) equipped at ISSP, the University of Tokyo. The ac measurements were performed with oscillating magnetic fields of $H_{ac}$ = 1 Oe at frequencies of 10, 100, and 1000 Hz. The temperature dependence of the heat capacity was measured using the conventional relaxation method in a physical property measurement system (PPMS; Quantum Design) at ISSP, the University of Tokyo. Magnetization curves up to approximately 50 T were measured by an induction method with a multilayer pulsed magnet at the International MegaGauss Science Laboratory of the Institute for Solid State Physics at the University of Tokyo.

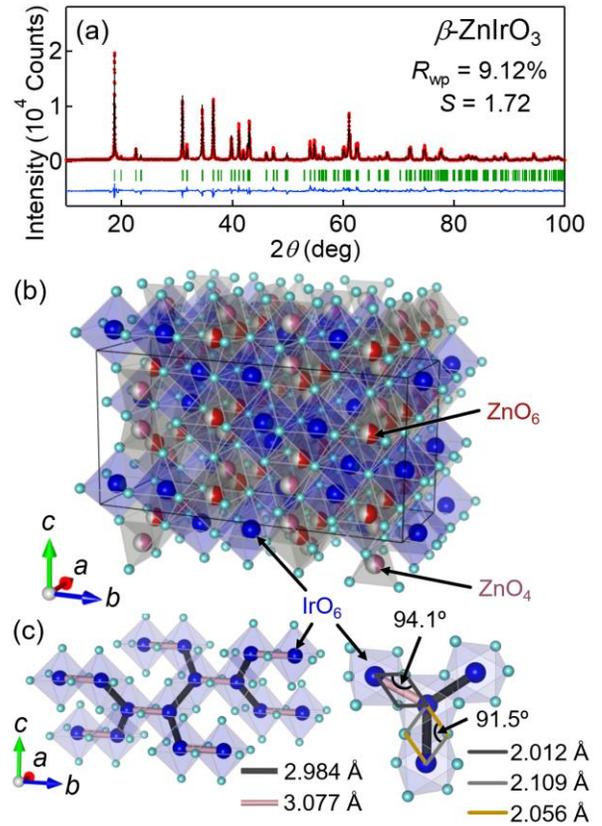

**Fig. 1** (**a**) Powder x-ray diffraction patterns of β-ZnIrO$_3$. The observed intensities (red circles), calculated intensities (black line), and their differences (blue curve at the bottom) are shown. Vertical bars indicate the positions of Bragg reflections. (**b**) Crystal structure of β-ZnIrO$_3$. (**c**) The left part shows the hyperhoneycomb network, where two Ir-Ir bonds with different lengths are shown differently. The right part shows the local lattice network of IrO$_6$ octahedra of β-ZnIrO$_3$ with two different superexchange pathways with different Ir-O-Ir angles. The VESTA program is used for visualization [36].

**Table 1.** Crystallographic parameters for β-ZnIrO$_3$ (Space group: *Fddd*) determined from powder x-ray diffraction experiments. The obtained lattice parameters are $a$ = 5.92705 (3) Å, $b$ = 8.76465 (3) Å, and $c$ = 17.8315 (3) Å. $B$ is the atomic displacement parameter. The occupancy factor $g$ of Zn1 and Zn2 is fixed to be summed to 1.

| atom | site | $g$ | $x$ | $y$ | $z$ | $B$ (Å$^2$) |
|---|---|---|---|---|---|---|
| Ir  | 16g | 1        | 1/8      | 1/8       | 0.3882(6) | 0.162(3) |
| O1  | 16e | 1        | 0.367(2) | 1/8       | 1/8       | 1.0(1)   |
| O2  | 32h | 1        | 0.141(1) | 0.3649(7) | 0.0462(5) | 1.0(1)   |
| Zn1 | 16g | 0.523(3) | 3/8      | 3/8       | 1/8       | 0.31(1)  |
| Zn2 | 16f | 0.477(3) | 3/8      | 1/2       | 3/8       | 0.31(1)  |

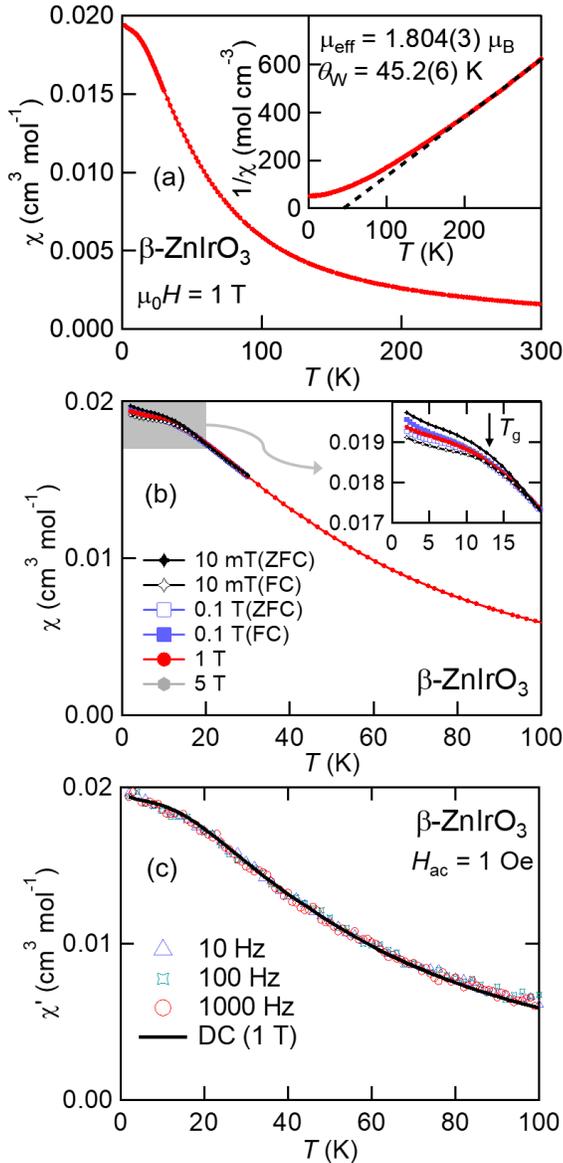

**Fig. 2** (a) Temperature dependence of magnetic susceptibility $\chi$ in β-ZnIrO$_3$ measured at an applied magnetic field of 1 T. The inset shows the reciprocal magnetic susceptibility $1/\chi$. The dashed line indicates the result of the Curie-Weiss fitting. (b) Temperature dependences of $\chi$ measured at several magnetic fields. The inset shows the enlarged view of shaded area. (c) The temperature dependences of the real part of the ac susceptibility measured in oscillating magnetic fields of $H_{ac}$ = 1 Oe at frequencies of 10, 100, and 1000 Hz.

*Results*. Figure 1(a) shows a powder XRD pattern of β-ZnIrO$_3$. All peaks can be indexed to an orthorhombic unit cell with $a$ = 5.92705 (3) Å, $b$ = 8.76465 (3) Å, and $c$ = 17.8315 (3) Å based on the space group of *Fddd*, which is the same of precursor β-Li$_2$IrO$_3$. The chemical composition was examined using energy dispersive x-ray spectrometry, yielding the ratio of Zn/Ir = 1.00(4). Thus, we concluded that the ion-exchange reaction is completed. The XRD pattern could not be reproduced by a model that the original Li$^+$ octahedral sites are replaced by Zn$^{2+}$ at half occupancy. Thus, we constructed an initial structure in which zinc ions occupy octahedral and tetrahedral sites. This Rietveld refinement converges well with the crystal structure shown in Fig. 1(b). Table 1 shows the crystallographic parameters. Zn$^{2+}$ ions occupy approximately half of the octahedral and half of the tetrahedral sites, respectively. Such a distribution of Zn ions is inevitably expected to generate randomness effects in the spin system.

There are two types of Ir-Ir bonds with different lengths, as shown in the left part of Fig. 1(c). One of the shorter Ir-Ir bonds of 2.984 Å forms a zig-zag chain (gray) linked by the longer Ir-Ir bonds of 3.077 Å (yellow) with alternating rotation by 68.13° around the *c*-axis. The difference between the two Ir-Ir bonds is ~3%, which is more anisotropic than the precursor β-Li$_2$IrO$_3$ (~0.2%) [28]. This non-equivalency of Ir-Ir bonds also leads to the non-equivalence of Ir-O-Ir bond angles. The two bond angles in β-Li$_2$IrO$_3$ are almost equivalent (94.5° and 94.7°), whereas those in β-ZnIrO$_3$ are more deviated (94.1° and 91.5°). The difference in the angle of the superexchange path should affect the magnetism.

Figure 2(a) shows the temperature dependence of magnetic susceptibility $\chi$ and its inverse $1/\chi$ measured at $\mu_0 H$ = 1 T. The Curie-Weiss (CW) fit of the $1/\chi$ data in the range of 200–300 K yields an effective magnetic moment $\mu_{eff}$ = 1.804(3) $\mu_B$ with a Weiss temperature $\theta_W$ = 45.2(6) K. The $\mu_{eff}$ value evidences a $J_{eff}$ = 1/2 pseudospin with a Lande's g-factor of $g$ = 2.08. This enhancement is due to the spin-orbit interaction. The positive $\theta_W$ value indicates predominantly ferromagnetic interactions among the $J_{eff}$ = 1/2 pseudospins, consistent with a Kitaev-type interaction. These parameters of β-ZnIrO$_3$ are similar but slightly larger than those of β-Li$_2$IrO$_3$ ($\mu_{eff}$ = 1.61 $\mu_B$ and $\theta_W$ = 40 K) [28]. In the low-temperature region of Fig. 2(a), the $1/\chi$ curve gradually deviates from the CW line below approximately 120 K, suggesting the development of short-range order (SRO). At lower temperatures, the $\chi$ data tend to saturate and approach a finite value with little magnetic field dependence, as shown in Fig. 2(b). At the lowest field of 10 mT, $\chi$ exhibits a slight thermal hysteresis between the zero-field-cooled and field-cooled data below $T_g$ = 12 K, suggesting a trace of spin glass-like contribution. The difference, however, is less than 4% of the total magnetization at 2 K, and the glassy component becomes negligible at high magnetic fields above 1 T. Moreover, Fig. 2(c) displays the real part of the ac susceptibility, which shows no frequency dependence without an anomaly and is completely consistent with the dc susceptibility. Furthermore, samples synthesized using ZnCl$_2$ instead of ZnSO$_4$ show a large spin glass-like contribution with a substantial sample dependence (see Supplemental Material [34]). Thus, the glass-like behavior would originate from frozen minority spins or a trace of magnetic impurities.

Fig 3(a) shows the temperature dependence of the heat capacity divided by temperature $C/T$ up to 250 K. There appears to be no anomaly indicating a long-range order not only at $T_g$ [see also the inset of Fig. 3(a)] but also in all measured temperature range. The $C/T$ versus $T^2$ plot at low temperatures is shown in Fig. 3(b), where $C/T$ exhibits linear dependence as a function of $T^2$ with a finite intercept of $\gamma$ = 5.31(7) mJ mol$^{-1}$ K$^{-2}$, which was obtained by fitting to the equation $C/T = \gamma + \beta T^2$. The $\gamma$-value is surprisingly independent of an applied magnetic field, strongly suggesting that the low energy spin excitation has nothing to do with the glassy contribution but derives from a QSL ground state.

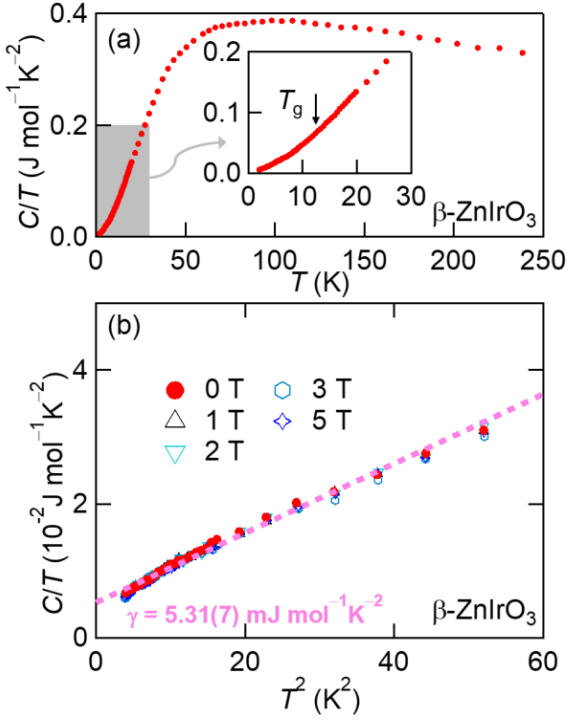

**Fig. 3** (a) Temperature dependence of the heat capacity divided by temperature $C/T$ for β-ZnIrO$_3$. The inset shows the enlarged view of shaded area. (b) $C/T$ versus $T^2$ plots under various magnetic fields up to 5 T at low-tempratures. The dashed pink line represents a fit to the equation $C/T = \gamma + \beta T^2$.

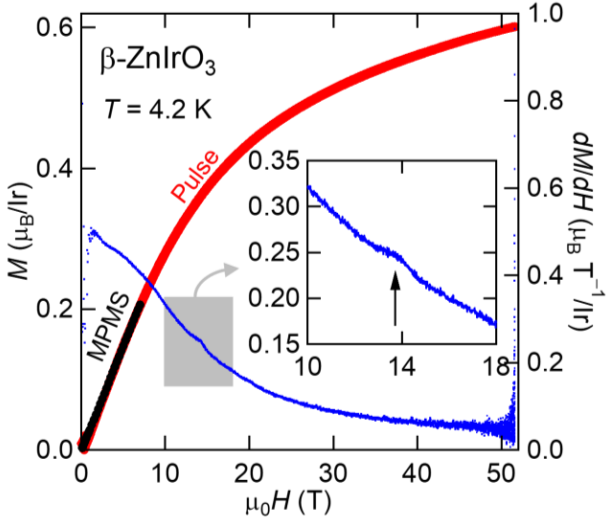

**Fig. 4** Magnetization curves $M$ and their derivative $dM/dH$ measured at 4.2 K for β-ZnIrO$_3$. The black circles were obtained in static fields below 7 T (MPMS), and the red circles were obtained at pulsed magnetic fields up to 51.5 T, calibrated to the former data. The arrow indicates an anomaly at ~14 T in the $dM/dH$ data. The inset shows an enlarged view of the $dM/dH$ data around the anomaly.

Magnetization measurements up to 51.5 T using pulsed magnetic fields were performed, as shown in Fig. 4. The magnetization $M$ shows convex-upward-increasing with a tendency to saturate. However, the $M$ value at 51.5 T is about 0.6 μ$_B$, much smaller than the saturated $M \sim 1$ μ$_B$ expected in the $J_{eff} = 1/2$ spin. In the theoretical predictions about the Kitaev QSL, $M$ eventually saturates at 1/2 μ$_B$ owing to a significant spin fluctuation in the Kitaev paramagnetic state [37]. Therefore, this saturating behavior in β-ZnIrO$_3$ may reflect the strong spin fluctuation. Moreover, the $M$ data shows a slight increase at ~14 T, evidenced by an anomaly in its derivative $dM/dH$. In the present stage, there is no decisive proof of a phase transition from Kitaev QSL to a paramagnetic state as theoretically predicted [38, 39].

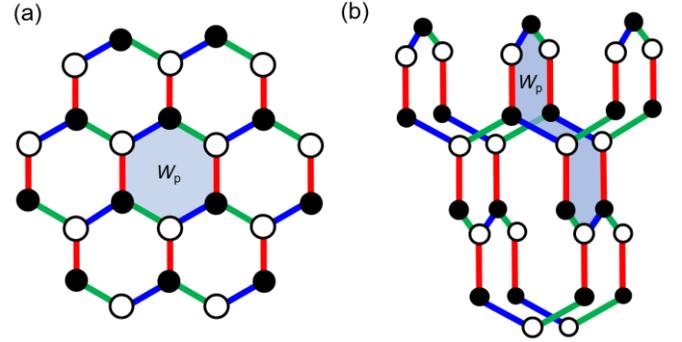

**Fig. 5** The Kitaev model on (a) 2D honeycomb and (b) 3D hyperhoneycomb lattice. The blue, green, and red colored bonds denote the exchange couplings $K_x$, $K_y$, and $K_z$ in the Kitaev Hamiltonian, respectively. The shaded $p$-sites closed-loop on each lattice represents the $Z_2$ flux denoted by the loop operator $W_p$ [1]; $p$ equals 6 and 10 for honeycomb and hyperhoneycomb lattice, respectively.

*Discussion.* As described above, β-ZnIrO$_3$ is a spin-orbit-coupled Mott insulator with $J_{eff} = 1/2$ hyperhoneycomb lattice. Signatures of QSL have been observed despite the dominance of ferromagnetic interactions, likely representing the Kitaev model. Many issues, however, remain and should be tackled urgently.

Firstly, randomness due to the irregular arrangement of Zn$^{2+}$ ions should influence magnetism. This randomness possibly causes spin-glass transition or generation of free-spins. However, no physical properties have been observed to characterize the disordered system. Moreover, the low-temperature heat capacity of β-ZnIrO$_3$ is almost independent of the magnetic field despite randomness. In contrast, the strong magnetic field dependence in $C/T$ and χ reproduced by scaling behavior has been observed in other Kitaev systems such as H$_3$LiIr$_2$O$_6$, Cu$_2$IrO$_3$, and α-Ru$_{0.8}$Ir$_{0.2}$Cl$_3$ [21, 40, 41]. The presence of quenched disorder theoretically explains well the scaling behavior of thermodynamic quantities [42]. These materials and β-ZnIrO$_3$ would be in the category of Kitaev QSL, but their magnetic dimensionality is different—2D honeycomb and 3D hyperhoneycomb. Our observation indicates that the Kitaev QSL on the 3D lattice is essentially less affected by bond randomness. In the excited state of

Kitaev QSL, the localized $Z_2$ fluxes are thermally excited in the form of closed loops [1,5-7]. Since the minimal *p*-sites closed-loop is different in 2D ($p = 6$) and 3D ($p = 10$), as shown in Figs. 5(a) and (b) [31,32], there would be a difference in the stability against randomness. As shown in Fig. 5(c), in the 2D Kitaev model, it is theoretically predicted that the effect of randomness will produce a larger flux with the 12-sites loop consisting of three hexagons [43]. The size effect of the flux as these excited states may be responsible for the difference in robustness to disturbance. Thus, it is plausible that the protective effect against randomness is hidden in the 3D Kitaev QSL, which will be clarified theoretically in the future.

The next question is whether or not the theoretically predicted double $C_{mag}$ peak exists. In the pure Kitaev model, spins are thermally fractionalized into itinerant Majorana fermions and localized fluxes, which release entropy in the high and low-temperature regions, respectively, which results in generating a double-peak structure at $T_H$ and $T_L$ in $C_{mag}$ [44]. However, in the theoretical calculation, the randomness effect on itinerant Majorana fermions and localized fluxes is different—randomness hardly interferes with the entropy release of itinerant Majorana fermions; on the contrary, it significantly suppresses the entropy release of localized fluxes [43]. Thus, $C_{mag}$ in β-ZnIrO$_3$ may be affected by randomness, resulting in the smearing of the entropy release of the localized fluxes. Besides, a finite-temperature "gas-liquid phase transition" to QSL is predicted in the limited model with strong anisotropy ($K_z >> K_x$, $K_y$ in the Kitaev Hamiltonian) in the hyperhoneycomb lattice [45]. Unfortunately, attempts to separate the magnetic and phonon contributions were unsuccessful, given the absence of a suitable phonon reference for β-ZnIrO$_3$. Therefore, the observation of fractionalization requires another probe that is not affected by phonon contributions. Furthermore, it has been predicted that the nuclear magnetic resonance (NMR) relaxation rate $1/T_1$ will show significant temperature changes at $T_H$ and $T_L$ [32]. Therefore, the NMR experiments on $^{17}$O-enriched β-ZnIrO$_3$ would be a promising method to observe the fractionalization; it is an issue for future work.

*Summary*. We have successfully synthesized a hyperhoneycomb lattice iridate β-ZnIrO$_3$ via a topochemical reaction. Although some signatures of Kitaev QSL have been observed, there are still many unresolved issues—some deviations from theoretical predictions can be identified. However, in contrast to the previous 2D Kitaev candidates, β-ZnIrO$_3$ shows certain robustness to randomness effects. Thus, although some problems are still to be solved, such as the protective effect against randomness, the present results likely suggest that β-ZnIrO$_3$ is the promising candidate for 3D Kitaev QSL.

The authors thank Z. Hiroi for various discussions. This work was supported by the Japan Society for the Promotion of Science (JSPS) KAKENHI Grant Number JP19K14646 and JP21K03441. Part of this work was carried out by the joint research in the Institute for Solid State Physics, the University of Tokyo.

# Supplemental material for "Quantum paramagnetism in the hyperhoneycomb Kitaev magnet β-ZnIrO$_3$."

Yuya Haraguchi[1,†], Akira Matsuo[2], Koichi Kindo[2], and Hiroko Aruga Katori[1,3]
[1]Department of Applied Physics and Chemical Engineering, Tokyo University of Agriculture and Technology, Koganei, Tokyo 184-8588, Japan
[2]The Institute for Solid State Physics, The University of Tokyo, Kashiwa, Chiba 277-8581, Japan
[3]Research Center for Thermal and Entropic Science, Graduate School of Science, Osaka University, Toyonaka, Osaka 560-0043, Japan
†chiyuya3@go.tuat.ac.jp

## Sample dependence of β-ZnIrO$_3$

Figure 1 in the supplementary information shows the temperature dependence of six different β-ZnIrO$_3$ samples. As discussed in the main text, the samples synthesized with ZnCl$_2$ showed strong sample dependence, while those synthesized with the mixed salt of ZnSO$_4$ and KCl showed almost no sample dependence. As shown in Figure 2 in the supplementary information, there is no particular sample dependence on heat capacity. As mentioned in the main text, the ZnCl$_2$ sample did not undergo a redox-neutral reaction, which may have changed the valence of Ir ions and caused a strong spin-glass due to electron or hole-doping. On the other hand, the trace amount of spin glassy contribution in the ZnSO$_4$ samples may also be due to a slight change in Ir-valence, as the complete redox-neutral reaction has not been realized. Hence, we can conclude that a trace amount of spin-glass contribution in the ZnSO$_4$ samples is not intrinsic.

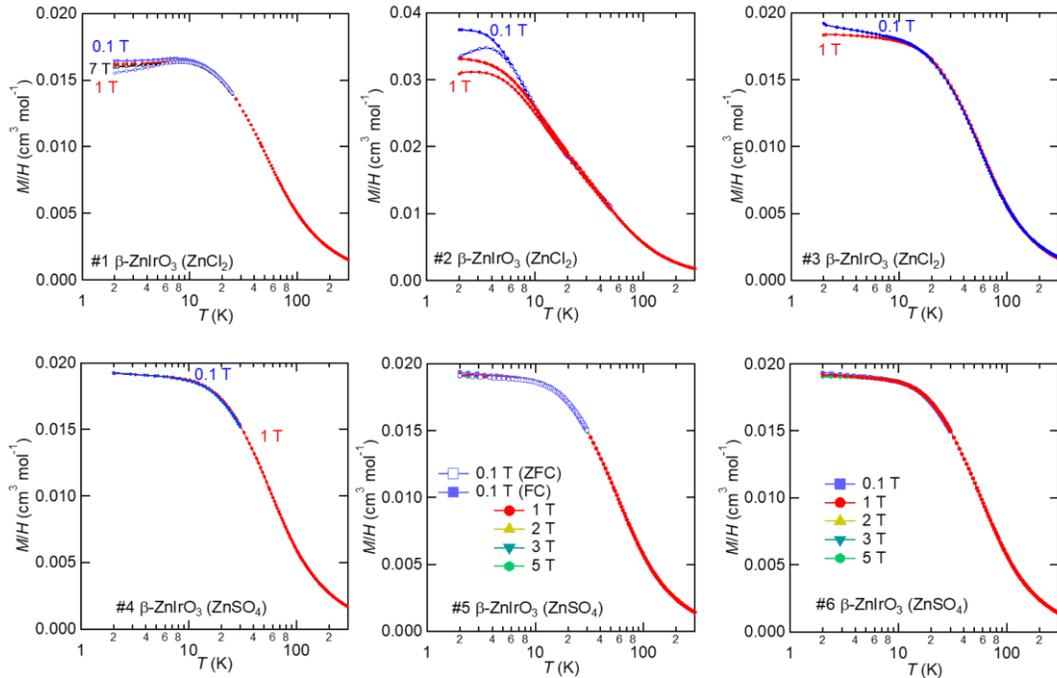

**Fig. 1.** Magnetic susceptibility of six different β-ZnIrO$_3$ samples. The three upper panels show the magnetic susceptibility of a sample ion-exchanged with ZnCl$_2$. The three lower panels show the magnetic susceptibility of a sample ion-exchanged with a mixture of ZnSO$_4$ and KCl salts. All data of XRD, magnetization and heat capacity in the main text is measured using the #5 sample.

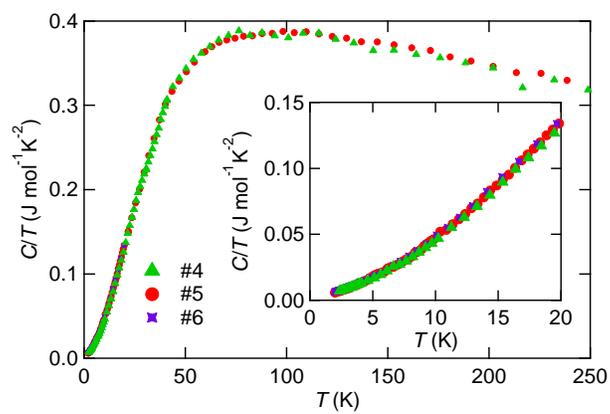

**Fig. 2.** Heat capacity of #4, #5, and #6 samples of β-ZnIrO$_3$. The inset shows the enlarged view at low temperatures.